# Towards density and phase space compression of molecular gases in magneto-electrostatic traps

Yuval Shagam and Edvardas Narevicius

Department of Chemical Physics, Weizmann Institute of Science, Rehovot 76100, Israel

**Abstract**
We introduce, analyze, and compare two novel methods of Single Photon Cooling that generically cool and compress molecular gases. The first method compresses the molecular gas density by three orders of magnitude and increases collision frequency in trapped samples. The second method compresses the phase space density of the gas by at least two orders of magnitude. Designed with combinations of electric and magnetic fields these methods cool the molecules from ~100mK to 1mK using a single irreversible state change. They can be regarded as generic cooling schemes applicable to any molecule with a magnetic and electric dipole moment. The high efficiency calculated, compared to schemes involving cycling, is a result of cooling the molecules with a single step.

**Introduction**
For many years the scientific community has been striving to develop general methods of molecular and atomic cooling for the fields of cold chemistry and physics. This ability would open the door to many different experimental categories such as the observation of chemical reactions in uncharted temperature regimes, observation of quantum degeneracy (BEC) in a wide variety of molecule and atom species and quantum computing [1]. Such a source would allow for a great increase in the precision of molecular spectroscopy and measurements of fundamental constants [2, 3]. In the cold temperature regime (below 1K) the deBroglie wavelength is of the order of the interaction potential where quantum effects should dominate, yet this regime is largely unmeasured even though many theoretical predictions exist [4-6]. Indeed the ultra-cold regime below 1mK, where quantum degenerate molecular gases can form, is even more mysterious and unexplored [7].

Sources of cold molecules (~$10^2$-$10^3$mK) have existed for many years such as cooling via supersonic beam expansion [8] or cooling via collisions (buffer-gas cooling) [9], yet observations of chemical reactions below 1K have been limited to charged or laser cooled species. Supersonic beam expansion creates cold beams (~100mK) which due to enthalpy conservation travel at supersonic velocities. Many methods to slow these beams have been explored such as Stark deceleration [10-12], Zeeman deceleration [13, 14], optical deceleration [15], and mechanical deceleration [16, 17]. However, the resulting particle density is too low to observe chemical reactions. Recently we have developed a co-moving magnetic trap which utilizes the Zeeman Effect in a 3-D moving magnetic trap to slow any type of particle that has a magnetic dipole moment without compromising the density or temperature, removing over 98% of the kinetic energy [18, 19].

The ultra-cold regime has been reached with a very specific set of atomic and molecular gases. The original method used to reach this regime was laser cooling which was developed over a decade and a half ago. When combined with evaporative cooling it can effectively cool specific atomic species to the nano-Kelvin range. Direct laser cooling on molecules is a difficult task because of the complexity of the molecular structure and the absence of lasers at the required wavelengths. It is possible to find specific molecules that have particularly favorable Franck-Condon transition factors such that direct laser cooling can be performed as was demonstrated by the Demille group with SrF [20]. Currently the most successful methods for molecular cooling to the ultra-cold regime are atomic based techniques in which atoms are cooled to the nano-Kelvin range and then combined via photo-association [21, 22] or Feschbach resonances [23, 24] to form molecules without heating. These methods create the coldest molecular gases to date, but are not general as they are limited to atomic precursors that are amenable to laser cooling. Other approaches include cycling the molecules between the vibrational and rotational states removing energy with every cycle in a Sisyphus type scheme as proposed by Pritchard [25]. The difficulty in such a method is finding a closed cycle within the complexity of the molecular structure by managing all of the allowed transitions. The Rempe group has demonstrated that this is feasible in an experimental implementation of such a cycling scheme, cooling a molecular gas of $CF_3H$ from 100mK to 1mK in about 10 seconds [26], which is of the order of magnitude of the life-time of the trap. The importance of reaching the 1mK threshold is that at this temperature loading of the gas into an optical dipole trap should be possible allowing further cooling by evaporative cooling. This is necessary since evaporative cooling in an electro-static and magneto-static trap is not possible due to state changing collisions [27]. Finally, an irreversible loading step can be used in order to accumulate molecules in a static trap as proposed and implemented by the Meijer group [28, 29].

The purpose of the methods introduced in this paper is to efficiently and generically cool a molecular sample, entrained in a supersonic beam and slowed by the co-moving magnetic trap decelerator, from ~100mK to any desired value above 1mK. These methods are tools for the exploration of this range and for the eventual loading of the gas into an optical dipole trap with higher efficiencies than those demonstrated to date. Both methods are based on Single Photon Cooling (SPC), which was introduced and implemented by the Raizen group for Rb atoms [30]. The Steck group has also implemented a Maxwell's Demon based method to cool atoms using a one way optical barrier [31], which was proposed by Raizen et al. [32]. The problem in direct application of SPC to molecules is that the polarizability of small radicals is usually too low and high laser power (~1kW) [33] is needed in order to create adequately deep optical dipole traps. Even with such high laser power the efficiency of SPC into an optical dipole trap would be low (~$10^{-8}$), due to the small phase space in comparison with the initial sample phase space density.

**Single Photon Cooling**
In Single Photon Cooling (SPC) particles are irreversibly transferred from one state to another around the particles' classical turning point, changing their effective potential surface. SPC takes advantage of the kinetic energy exchange with the conservative potential energy to lower the particle's total energy. The kinetic energy of such a particle will remain the same while it is trapped in a different potential surface according to the new state of the particle. It will be shown that a large tunable fraction of the total energy can be removed with such a process. The OH radical in the $^2\Pi_{3/2}$ electronic ground state will be used as a practical example. The total angular momentum of the OH radical in the ground state is J=3/2. There are four Zeeman sublevels with different projections of the total angular momentum on the quantization axis, as displayed in Fig 1.

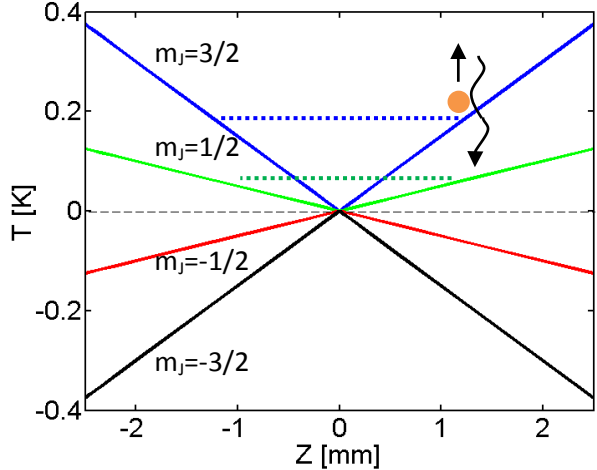

Fig. 1: Adiabatic energy levels of the OH radical in the $^2\Pi_{3/2}$ electronic ground state in a magnetic quadrupole potential. In SPC a molecule is excited around its classical turning point at the fully stretched $m_J=3/2$ and may decay to $m_J=1/2$ decreasing its energy 3 fold.

Assume that the process begins with a cloud of cold OH radicals in the fully stretched $m_J=3/2$ state. Suppose now that we irreversibly transfer a "hot" molecule at the classical turning point from the $m_J=3/2$ into the $m_J=1/2$ Zeeman sublevel (see Fig. 1). This process reduces the total energy of the molecule by a factor of 3, given by the slope ratio between the two potential surfaces. In order to remove a substantial amount of energy one needs to repeat this cycle many times, greatly increasing the time-scale of the process and lowering its efficiency. Clearly the molecular structure, which includes vibrational, rotational hyperfine energy levels complicates the possibility of cycling.

**SPC Utilizing a Magneto-Electrostatic Trap (Method I)**
Our idea is to use the molecular complexity to our advantage. We will show that by using combined Zeeman and Stark effects we can freely tune the slope ratio between two trapping potential surfaces and increase the cooling efficiency. After we trap OH radicals in a static magnetic quadrupole trap we superimpose an electrostatic quadrupole trap on top of the magnetic one. It is important that the electrical and magnetic fields are parallel in the trapping region, which allows the calculation of total energy as a sum of Stark and Zeeman shifts [34]. In Fig. 2a the contribution of the magnetic field to the energy levels is presented. The four Zeeman sublevels are easily recognizable. Let us analyze how these will be affected by the presence of electrical fields. We focus our attention on the low field seeking Stark levels from the $f$ doublet manifold that experience a linear and positive Stark effect. In Fig. 2b the contribution presented is due to the electric field alone. Since all four levels are from the $f$ doublet all four states are low field seekers (with respect to the electric field) with positive slopes only. In a combined magneto-electrostatic quadrupole trap all four levels will experience an additional positive energy shift. As a result, the slopes of the two sublevels, $|m_J=3/2, f\rangle$ and $|m_J=1/2, f\rangle$ will steepen, and the potential gradients of the two magnetically high field seeking states, $|m_J=-1/2, f\rangle$ and $|m_J=-3/2, f\rangle$ will also rise with the increasing electric field gradient. Whenever the contribution of the magnetic field is equal to that of the electrical field, the combined potential gradient of these two states is zero. Therefore at higher electrical field gradients the molecules in the $|m_J=-1/2, f\rangle$ state become confined in a very shallow potential well. The potential slope ratio between the $|m_J=3/2, f\rangle$ and $|m_J=-1/2, f\rangle$ states can be tuned to almost any value as a function of electric and magnetic field gradient ratio [35]. For example let us choose the magnetic and electrical fields such that a ratio of 1/70 in potential slopes is formed. We define this ratio as ε.

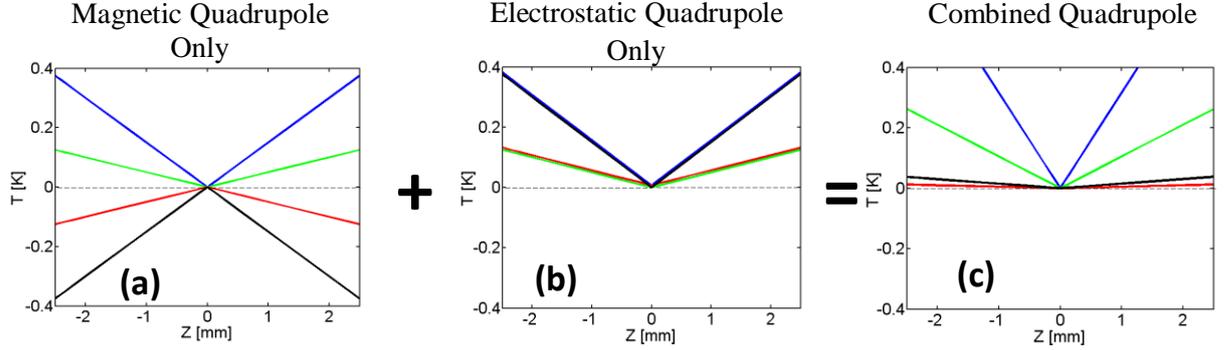

Fig. 2(a, b, c): Adiabatic energy levels of the OH radical in the $^2\Pi_{3/2}$ state in a magnetic quadrupole field (a), an electric quadruple (b) and a combination of the two in which the field lines are parallel (c).

The irreversible transition takes place near the classical turning point and changes the projection of the total angular momentum from 3/2 to -1/2, which will reduce the total energy of a molecule 70 (1/ε) fold in a single step, eliminating the need for cycling. This irreversible transition can easily be done by exciting the molecules on the Zeeman sensitive transition along the $R_2$ line of $A^2\Sigma_{1/2}$ (v=0) ← $X^2\Pi_{3/2}$ (v=0) transition at 308nm. We estimate that at least 15% (henceforth referred to as γ) of the molecules will decay to the |$m_J$=-1/2, $f$⟩ state. Since the transition is Zeeman sensitive, the transition location can be chosen by tuning the wavelength to match the energy difference at a certain point. Starting with the "hottest" molecules and slowly ramping the transition frequency down, lower energy molecules are accumulated in the shallow potential well, until the entire volume of the trap is swept.

In a spherically symmetric trap the spatial angular momentum and energy are conserved, which implies that the radius of the classical turning point (CTP) is constant. Since an excitation only removes translational energy in the radial direction, the molecules are left with residual translational energy in the angular direction. In the anisotropic case the angular momentum is no longer conserved. If the angular momentum of a molecule varies quickly, compared to the time-scale of the sweep process, then its CTP also changes rapidly. When the CTP moves up the potential barrier of the trap, energy conservation dictates that the residual angular velocity will be lower. This means that the efficiency and density compression would only improve, since SPC "catches" the molecules higher up the potential barrier of the trap, due to the sweep order of the excitation process. Thus, modeling and simulating the process in a spherically symmetric potential, as opposed to an anisotropic potential, provides a lower limit for these values. Therefore we estimate the portion of the molecules that can be trapped by energy considerations is ε [36], which corresponds to energy removal in one dimension. The resulting loading efficiency into the shallow trap is:

$$\eta = \gamma\varepsilon \quad Eq.(1)$$

We estimate that for the case 1/ε = 70 and γ = 15% the loading efficiency would be greater than $10^{-3}$. We begin with a supersonic beam at ~100mK. As we have shown our moving magnetic trap decelerator conserves number density down to stopping velocities so we expect to trap at least $10^9$ OH radicals in a 5x5x5 mm$^3$ volume trap (a density of ~$10^{10}$ molecules/cm$^3$), in the fully stretched state [18]. Consequently, at the end of the process we will have $10^6$ molecules trapped in a 5x5x5 mm$^3$ trap at a temperature of 2mK. Now, as a final stage, the cold cloud can be adiabatically compressed (conserving the phase space density) by increasing the electrical field gradient. The trap size can be reduced by a factor of 70 (1/ε) in every dimension (linear potential), such that the trap volume is reduced by a factor of ~$10^5$. The density of the trapped OH radicals will be $10^{12}$ molecules/cm$^3$ at the end of the process, four orders of magnitude higher than densities reported to date. At this density the collision rate is about 1kHz and should easily be observable. More generally adiabatically resizing the trap compresses the molecule density as shown by the following equation:

$$C_\rho = \frac{\gamma\varepsilon}{\varepsilon^3} = \gamma\varepsilon^{-2} \quad Eq.(2)$$

Monte Carlo simulations of this process were performed in order to verify the order of magnitude of the efficiency, as well as the density compression predicted by this rough model. The results obtained in the simulations (Fig. 3) indeed verify the model.

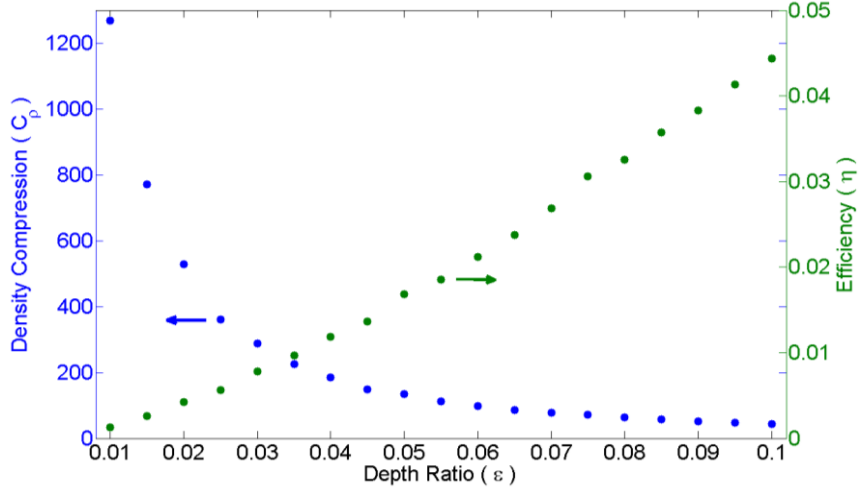

Fig. 3: Monte Carlo simulations of method I over 216ms. The density compression and efficiency are plotted as a function of the trap depth ratio ε. The simulations were performed with an initial uniform distribution in space and velocity, and γ was taken as 15%. The results are in line with the orders of magnitude predicted by the model (Eq. 1, 2).

In the simulations a uniform distribution of molecules, both in velocity and position, was randomly placed in the trap such that a constant number remained trapped inside. The trap potential used was linear in the radial direction as is formed by a magnetic quadrupole, and perfectly spherically symmetric. The previous discussion has shown that such an isotropic potential would provide the minimal results in terms of efficiency and compression.

The reason a uniform distribution is used is due to the simplistic mathematical models developed, which uniformly select parts of the distribution at different stages in the schemes. These mathematical models are only designed to indicate the order of magnitude of the physical properties of the scheme. In fact taking Gaussian initial distributions only further improved the simulation results, though not by an order of magnitude, which is easily understood since the trapped molecules have lower energies than in the uniform case. The uniform distribution makes the ratio ε the only physical aspect needed to describe the potentials.

The molecules were propagated in space-time via the Runge-Kutta numerical approximation. The irreversible state change was created via an instantaneous excitation to all molecules in a region (replicating a saturated transition) according to the scheme. Of them 15% (γ) selected at random decay to the state |$m_J$=-1/2, $f$⟩, retaining their position and velocity, and the rest are removed from the simulation.

The region of excitation was a shell with a width $1/40^{th}$ of that of the radius of the trap. All shells underwent the excitation process sequentially from the outer portion of the trap to the inner portion. Each shell was excited for a period of 3.6ms for a total of 144ms for the entire excitation process. The molecules were then propagated in the shallower well, whose slope is determined by $\varepsilon$, without excitation for another 62ms to allow the over energetic ones to escape. This final period is important to verify the number of trapped molecules.

The main advantage of this method is its efficiency, which should be high enough to load molecules into a shallow optical dipole trap to continue further cooling and compression of the phase space density (PSD). However, the main drawback of this method is that it does not compress the PSD (as PSD stays constant during adiabatic resizing of the trap), but can only serve as a stepping stone to evaporative cooling (which will compress the PSD). The second method does compress the PSD as we will show in the next section.

**SPC in a Magnetic Trap with Electric Field Barrier (Method II)**
In method II we again utilize both the Zeeman and Stark effects. We begin with a magnetic quadrupole trap, such that the energy levels of the OH radical in the $^2\Pi_{3/2}$ electronic ground state are as illustrated in Fig. 1. 'Trap A' is formed for radicals in the state $|m_J=3/2, f\rangle$ as shown in Figure 4a. Utilizing the Stark effect we add an electric field gradient near the classical turning point of the "hottest" molecules in 'Trap A', which can be seen by the perturbation in the energy levels for values of Z above 3mm displayed in Fig. 4a. We will focus our attention on the low field seeking Stark levels from the $f$ doublet manifold. The electric field gradient will act as a barrier for molecules with parity $f$, such that it blocks movement in the $\hat{Z}$ direction and confines them in the transverse direction.

A possible implementation of such a barrier, as shown in Fig. 4b, is an electrostatic trap without the electrode that closes the entrance. This configuration consists of two ring shaped electrodes with different radii and opposite potentials. Both the low field seeking and high field seeking molecules, with respect to the magnetic field, will see the electric field gradient as a barrier. For molecules in the $|m_J=-1/2, f\rangle$ state the electric field gradient combines with the magnetic field gradient, which blocks the entrance, forming a shallower and spatially smaller trap off-center from 'Trap A'. We will refer to this trap as 'Trap B'.

The Stark and Zeeman adiabatic energy levels are simply summed without regarding the angle between the electric and magnetic fields, since the electric field is small enough (<1kV/cm). In this regime both the Stark and Zeeman shifts are linear and there are no level crossings [37]. (In this scheme only the states with total angular momentum projections of 3/2 and -1/2 come into play.)

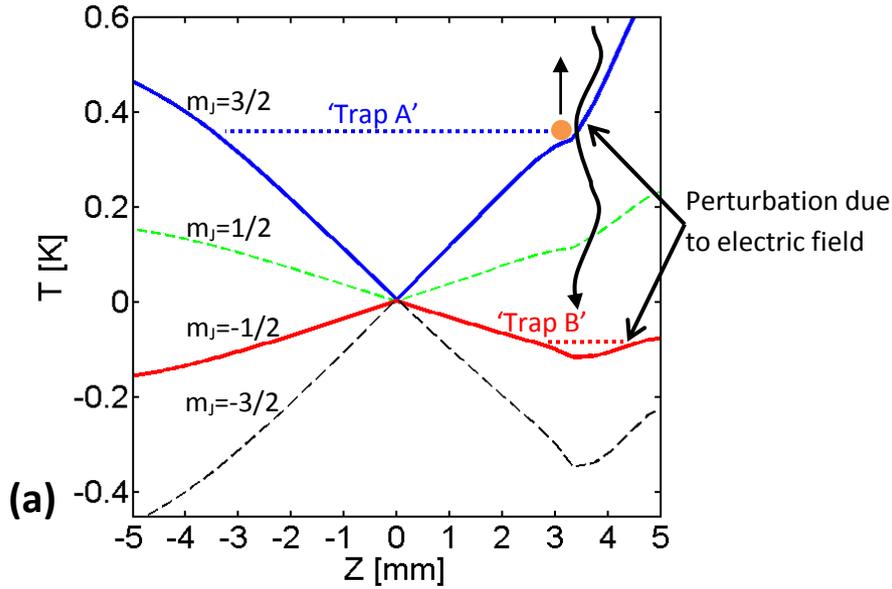

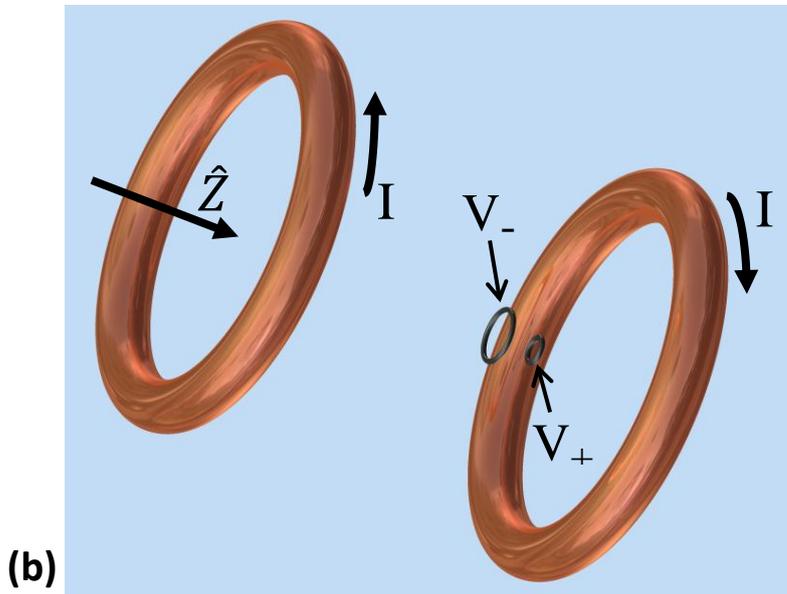

Fig. 4 (a, b): Figure (a) shows the adiabatic energy levels, of the combined Zeeman and Stark effects, which form 'Trap A' (Blue) and 'Trap B' (Red) on the Z axis. The potential is formed with a magnetic quadrupole formed by two current loops in an anti-Helmholtz configuration and half an electrostatic trap formed by two ring shaped electrodes with potentials of opposite polarity (b). This implementation creates the potential in (a) with confinement in the transverse direction, such that 'Trap B' is formed. The dotted energy levels in (a) are ignored in this scheme.

The molecules have a total angular momentum projection of 3/2 in their initial state. A laser focused on the position of 'Trap B' (near the "hottest" molecules' classical turning point) will create the same electronic excitation described in method I, which results in an estimated 15% (henceforth referred to as γ) of the molecules decaying to the $|m_J=-1/2,f\rangle$ state. In this state the molecules are trapped inside 'Trap B'. A magnetic bias can now be gradually applied, moving the classical turning point of molecules with lower kinetic energy towards the loading point, where they are also loaded into the trap. Alternatively the barrier can be mechanically moved along the Z axis. The depth and size of 'Trap B' are determined by the height of the electric barrier and the gradient of the magnetic field.

In this case the SPC does not take place over the entire volume of the large trap but only a portion with the volume of 'Trap B', differing from method I. Defining the ratio of the radius of 'Trap B' to the radius of 'Trap A' as $\chi$, we find that the instantaneous part of the spatial molecule distribution of 'Trap A' affected by SPC is $\sim\chi^3$. Taking into account the scanning of the entire Z coordinate via the tunable magnetic bias at least $\sim\chi^2$ of the distribution is affected by SPC. This is the minimal portion since the molecule locations are dynamic, and therefore the number of molecules that pass through the excitation volume grows in time. Therefore the spatial portion, which lies between $\sim\chi^2$ and 1, grows with the time-scale of the process.

We shall now analyze the efficiency in terms of energy. We must take notice that since the potential of 'Trap A' is spherically symmetric this cooling scheme one dimensional just like in Method I, where this deduction was discussed in detail. Therefore the portion of the molecules that can be trapped by energy considerations is $\varepsilon$ [36], where $\varepsilon$ is defined as the ratio of the energy depth of 'Trap B' to 'Trap A'. The resulting total minimum loading efficiency into 'Trap B', and minimal phase space compression are:

$$\eta_{\min} = \gamma\chi^2\varepsilon \quad Eq.(3)$$

$$C_{\min} = \frac{\gamma\chi^2\varepsilon}{\chi^3\varepsilon^{3/2}} = \gamma\chi^{-1}\varepsilon^{-1/2} \quad Eq.(4)$$

For the experimentally feasible ratios $\varepsilon=1/70$ and $\chi=1/10$ we find that $C_{\min}= 12.5$, compressing the phase space by at least 1 order of magnitude. Monte Carlo simulations of this case show that on a typical time-scale of the process (108ms) the resulting phase space compression was greater than 2 orders of magnitude! This strengthens the prediction above that the time-scale of the process can greatly enhance the performance with respect to the model discussed in Eq. 3 and Eq. 4. The simulation results are provided in Fig. 5. Indeed, the same simulation over much shorter time scales displayed results closer to the minimal results predicted by the model verifying the hypothesis proposed, that the difference between the model and the simulation is a result of the time-scale.

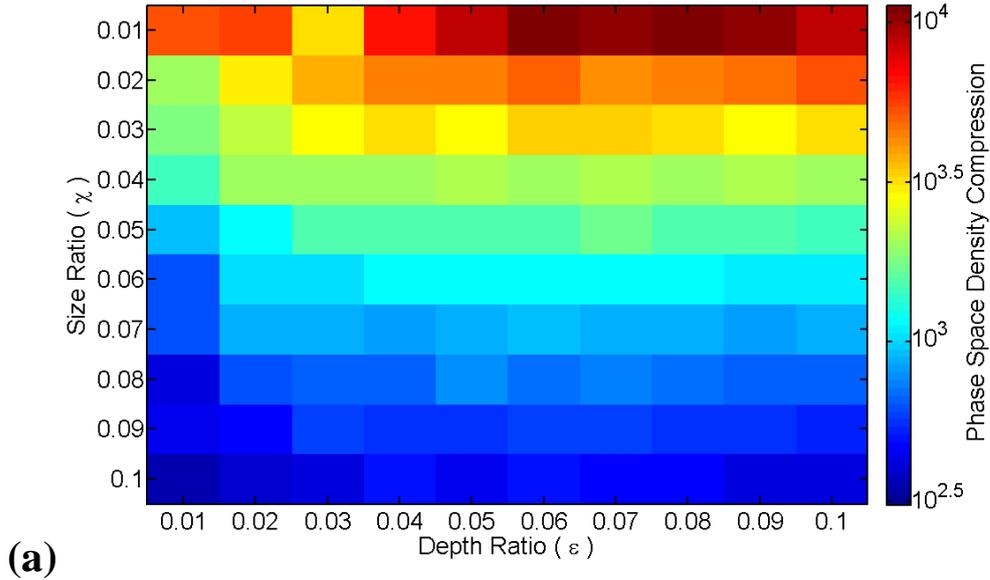

(a)

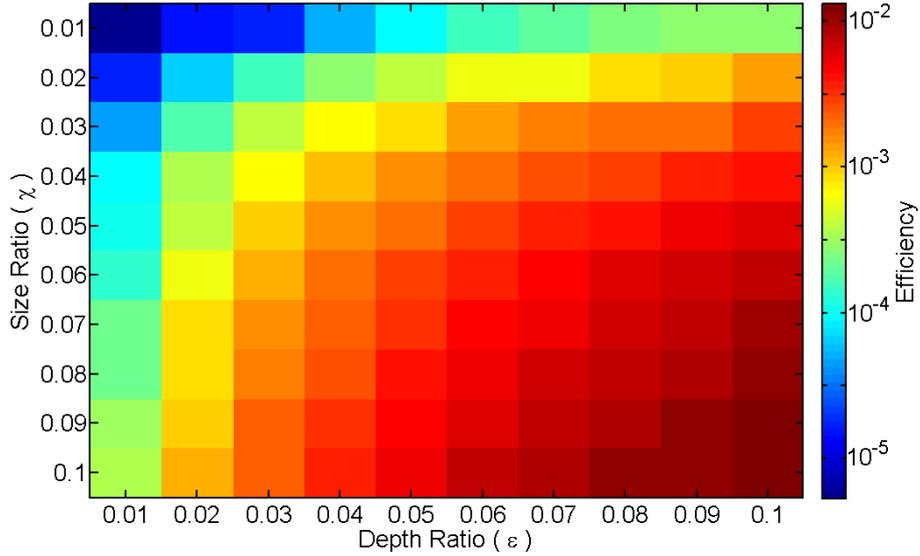

**(b)**

Fig. 5 (a, b): Phase space density compression (a) and efficiency (b) results of the Monte Carlo simulations of method II as a function of the spatial size ratio ($\chi$) and the energy depth ratio ($\varepsilon$). The process was simulated on a time-scale of 108ms with an initial uniform distribution in space and velocity. It is evident that the phase space compression results (a) are much better than the ones predicted in the model (Eq. 4). The best compression is attained for the smallest $\chi$, however these values may be difficult to implement in an experiment, and to detect due to low efficiency.

The description of the simulations is identical to the description provided in method I except for the following differences. In this case the uniform distribution makes the ratios $\varepsilon$ and $\chi$ the only physical aspects needed to describe the potentials. The region of excitation was a sphere with a radius determined by the proportion $\chi$. Molecules that decay to the state $|m_J=-1/2, f\rangle$ in the sphere were checked against the depth of the sphere, given by $\varepsilon$, according to their kinetic energy, and the over energetic ones were removed from the simulation. The sphere was moved continuously towards the center of the 'Trap A' ending the process when its edge reached the center. The entire process was simulated over 108ms.

Fig. 5a shows that the phase space density compression increases, as $\chi$ gets smaller. This effect is countered by the efficiency (Fig. 5b), which decreases, as $\chi$ gets smaller, lowering the total number of molecules trapped. Eventually, even though the compression increases it will become increasingly difficult to detect the resultant molecule samples, limiting the value of $\chi$. The number of molecules in the sample at the beginning of the process along with the signal to noise ratios of the detection techniques will dictate this lower limit, and subsequently the optimal choice of parameters.

**Conclusion**

The methods of implementation of SPC discussed and analyzed in this article are very general and can be modified to cool almost any molecule that has an electric and magnetic dipole moment. Method I excels in efficiency, while method II excels in phase space density compression. The time-scales spanned by these methods are two orders of magnitude lower than the total lifetime of the trap as opposed to any proposed cycling scheme. Both these advantages make these methods excellent candidates for stepping-stones for evaporative cooling, and for creating dense molecular gases at any temperature between 150mK and 1mK. Experiments will most likely show that specific molecules will work better with one of the methods according to their traits (i.e. initial densities, length of process time-scale possible, etc.). The robustness of these methods is a direct result of the lack of cycling. We plan to implement both of the methods and experimentally test the conclusions of this paper in the coming future.

**Acknowledgments**

This research is made possible in part by the historic generosity of the Harold Perlman Family. E.N. acknowledges support from the Israel Science Foundation.